# HUMAN-CETRIC TELEROBOTICS: INVESTIGATING USERS' PERFORMANCE AND WORKLOAD VIA VR-BASED EYE-TRACKING MEASURES

**Authors:** Federica Nenna, Davide Zanardi, Luciano Gamberini

**Abstract**. Virtual Reality (VR) is gaining ground in the robotics and teleoperation industry, opening new prospects as a novel computerized methodology to make humans interact with robots. In contrast with more conventional button-based teleoperations, VR allows users to use their physical movements to drive robotic systems in the virtual environment. The latest VR devices are also equipped with integrated eye-tracking, which constitutes an exceptional opportunity for monitoring users' workload online. However, such devices are fairly recent, and human factors have been consistently marginalized so far in telerobotics research. We thus covered these aspects by analyzing extensive behavioral data generated by 24 participants driving a simulated industrial robot in VR through a pick-and-place task. Users drove the robot via button-based and action-based controls and under low (single-task) and high (dual-task) mental demands. We collected self-reports, performance and eye-tracking data. Specifically, we asked i) how the interactive features of VR affect users' performance and workload, and additionally tested ii) the sensibility of diverse eye parameters in monitoring users' vigilance and workload throughout the task. Users performed faster and more accurately, while also showing a lower mental workload, when using an action-based VR control. Among the eye parameters, pupil size was the most resilient indicator of workload, as it was highly correlated with the self-reports and was not affected by the user's degree of physical motion in VR. Our results thus bring a fresh human-centric overview of human-robot interactions in VR, and systematically demonstrate the potential of VR devices for monitoring human factors in telerobotics contexts.

## 1. INTRODUCTION

Recently, the industrial sector has witnessed a massive shift of general interest from the machine to the human, who have become the core of the current industrial evolution. The manifest of Industry 4.0 is indeed the 'human-centric manufacturing' (Lu et al., 2022), which is placing the worker's wellbeing at the center of the production process. Such framework was proposed in response to the technical advancements in smart and intelligent manufacturing, which generated new challenges for the human operator. One practical example is provided by human-robot collaborative systems, which are also called collaborative robots or cobots (Krüger et al., 2009, Wang et al., 2017). This technology, child of Industry 4.0, removes the perceived boundaries of a manufacturing environment allowing humans to interact with machines in a fluid-tight manner, also remotely. The close integration of technology, robots, automated factory lines and intelligent manufacturing increased the skill demands for the operator, who now must cope with cognitive loads to work efficiently (Doolani et al., 2020).

Therefore, the proposed intention of Industry 4.0 to better understand the single worker in the field, with his/her individual needs and capabilities, now became an urge. Nonetheless, while the aims of a human-centric manufacturing are clear, practical applications are still scarce. Indeed, even the latest research works on industrial manufacturing demonstrate how higher attention is always dedicated to the technical aspects of industrial systems, while the analysis of human and cognitive factors relative to the worker is consistently neglected (Grandi et al., 2020; Smith and Sepasgozar, 2022).

Some researchers started to propose and test different approaches for assessing human factors such as user experience, stress, fatigue and mental workload while interacting with robotic systems (e.g., Chien et al., 2018; Villani et al., 2020). Furthermore, recent methodological advances support multimodal assessments, combining different behavioral and physiological tools to fully understand humans and their workload in the field (Dehais et al., 2020; Matthews et al., 2015). However, it is to acknowledge that measuring workload in highly interactive environments is not that easy (Dehais et al., 2020). Workload is commonly quantified via self-reports (e.g., NASA-TLX questionnaire, Hart and Staveland, 1988), which show diverse disadvantages: their administration inevitably disrupts the task flow, it is prone to inter-subject variability, and to the individual's ability to self-assess (Carswell et al., 2005). Given those limitations, the question arises whether and how workload can be assessed continuously, without task disruptions, and in a more objective way. Eye-tracking systems can help fill this gap, as they are increasingly portable and affordable and can capture workload-related eye behavior in the field (Novak et al., 2015). Also, the most recent Virtual Reality (VR) headsets are provided with an integrated eye-tracker that allows the measurement of eye-related physiological indexes within virtual environments, without disrupting users' actions. This is an exceptional opportunity for continuous workload monitoring during robotic teleoperation.

Not without reason, the present work focuses specifically on VR-based robotic teleoperations, which are gaining particular traction in the robotic teleoperation industry (Franzluebbers and Johnsen, 2019; Martin-Barrio et al., 2020; Rosen et al., 2018). As compared to traditional teleoperation means, VR allows to immerse in realistic environments and use interaction modalities that go far beyond the keyboard button press. This makes it possible for example to teleoperate a robot by using gestures or physical actions (e.g., Martin Barrio et al., 2019) rather than using the conventional keyboard, mouse and joysticks (e.g., You and Hauser, 2012; Mavridis et al., 2015), thus reproducing the same manual features of cobots. The strength of action-based controls is that they leverage natural and embodied controls, allowing users to directly manipulate a replica of the robot in a button-free fashion, and to perform physical actions or gestures similar to those they would perform if manipulating the physical robot itself. Such feature would streamline the amount of time typically spent on the robot programming significantly. On the other hand, it is also true that button-based controls are commercially widespread and thus more familiar to the most. Given the fairly novelty of VR-based robotic teleoperation, how the system control affects work efficiency and human factors (i.e., performance, workload) is not clear yet.

While being a novel computerized methodology in industrial engineering and robotic teleoperation, VR is also an exceptionally valuable tool for assessing human factors in simulated industry too. It has been demonstrated how VR allows to gain information on the qualitative experience while interacting with a product (e.g., Rebelo et al., 2012). Furthermore, extensive datalog can be obtained throughout any VR-based experience or operations. For instance, time series data on the position and rotation of the VR headset and controllers, along with time stamps of any interaction between the physical hardware and a virtual object, and also between virtual objects themselves, are continuously generated by the VR device and can be leveraged either online or offline. One of the most known software allowing to program virtual environments and managing data streams is Unity, which was also leveraged in simulated industrial robotics (Crespo et al., 2015; Naranjo et al., 2020; Nenna et al., 2022). All data generated by the interactions between the user and the virtual environment can inform both on the system state (exact position and rotation of a robotic arm) and, if properly processed, also on the human state. For instance, it is possible to automatically compute the level of efficiency of a human during a given task by using task time and accuracy, and deduct his/her levels of fatigue from the current work rhythms. Furthermore, with the latest eye-tracking-equipped VR devices, it is even possible to monitor workers' eye parameters such as pupil size, eye openness and blinks, which are known to change with workloads (Novak et al., 2015; Nenna et al., 2022). Such information can be extremely useful in view of a VR-based human-centric telerobotics, as they would allow to continuously monitor both human's and robot's activities.

The present work thus aims to: 1) investigate how the degree of interactivity of a control system affects the user's performance and workload when guiding an industrial robotic arm in VR; 2) leverage the VR-embedded eye-tracker and relate eye parameters continuously collected throughout the task with the traditionally self-reported workload. For these purposes, we conducted a systematic investigation on users driving a virtual replica of the robotic arm e-Series UR5e through a simple pick-and-place task, which simulated robotic teleoperation. We asked participants to perform the same task using different control systems (button-based and action-based) and under different levels of mental demand (single-task, dual-task). Performance was assessed via operation time and error rate at the task, while the human mental workload was assessed through both traditional self-reports and eye-tracking indexes detected from the VR headset. Our research questions and hypotheses are outlined as follows:

a) **Task load**. The dual-task paradigm employed in this work was previously tested with a physical and a virtual replica of the industrial robot e-Series UR5e (Nenna et al., 2022). As opposed to the authors' original paradigm, we here transposed the secondary arithmetic task in the visual domain to increase the task difficulty in the dual-task condition: according to resource theories, dual-task costs are particularly evident when multiple concurrent tasks share the same resources (Pashler, 1984). Therefore, participants needed to visually pay attention to the numbers to sum at the arithmetic task while also being engaged in precise visuomotor coordination for executing the pick-and-place task. As a methodological control, we thus expect the dual-task to affect participants' performance (slower operation times and higher error rates) and workload (higher NASA-TLX score, higher pupil size variation, lower perclos, and shorter and fewer blinks).
b) **Control system**. The core of our work revolves around the following research question: how the degree of interactivity of a control system affects users' performance and workload in virtual teleoperations? In contrast with the widespread button-based controls, action-based controls fully emphasize the opportunities offered by an interactive technology such as VR, and basically reproduce the same manual manipulation feature of collaborative robots in a virtual workspace. Therefore, by leveraging the use of natural and embodied controls, we hypothesize that an action-based system potentially represents a more efficient, intuitive, and less demanding solution for guiding robots compared to a button-based one, even in VR.

c) **Sensitivity of VR-embedded eye-tracker metrics to workload**. Eye-trackers-equipped VR headsets have the potential of capturing workload-related fluctuations in eye parameters while freely acting in the virtual environment. In this view, a relevant question is which of the eye parameters collected via the VR headset is the most sensitive to workload changes, also considering the higher degree of motion during VR interactions. To address this question, we inspected the relations between the self-reported workload and each of the investigated eye parameters (pupil size, PERCLOS, blink frequency and duration) both when using the action-based and button-based controls. We assume higher correlations for those eye parameters that are more sensitive to workload.

# 2. STATE OF THE ART

## 2.1 Comparative literature on control systems for robot teleoperation

Robotic systems can be teleoperated via various control systems, allowing a higher or lower degree of interactivity, which we here refer to respectively as Low Interactivity Control Systems (LICS) and High Interactivity Control Systems (HICS). LICSs include keyboard, mouse and joysticks (You and Hauser, 2012; Lu and et al., 2008), which are the most spread and familiar to the most. Differently, all those control systems allowing physical and often direct interactions with a machine are here considered as HICSs. For instance, by using motion capture technology, it is possible to leverage human gestures as a control system to guide mobile robots: users could physically indicate with their hand the position where the robot will move (Cicirelli et al., 2015). Similarly, it is possible to manipulate robots in VR through physical and direct interactions (e.g., Martin Barrio et al., 2019). A different case is a Master-Slave framework, which consists of controlling a teleoperated robot (slave) through the direct manual manipulation of a second robot (master) (e.g., Vozar, 2013). Further examples of HICSs involve kinesthetic or 3D Haptic devices that provide force feedback to the users (Martin and Hillier, 2009), which allow the manipulation of 3D objects in virtual environments and are thus particularly useful in teleoperation tasks (Berkley, 2003).

Among the studies that directly assessed the effects of different degrees of teleoperation system interactivity on users' performance, many evidenced the advantage of HICSs over LICSs. For example, Vozar (2013) tested a master-slave in contrast to joystick teleoperation. Participants drove a customized skid-steer robot through a delimited space to reach some boxes and then teleoperated the robot's arm to grab and move the boxes. Results revealed that teleoperating the robot via master-slave significantly improved performance when moving the boxes, it reduced the total time of the task and was rated as more intuitive and easier to use compared to the joystick. Gliesche and colleagues (2020) compared teleoperation performance in a sample of nurses using a haptic device or a keyboard and mouse for guiding a 7-degree-of-freedom robot manipulator through a desktop-based pick-and-place task. When using the haptic device rather than mouse and keyboard, participants showed shorter task completion times. A few studies also demonstrated how HICSs outperformed over LICSs in VR. For example, Franzluebbers and Johnsen (2019) assessed the performance of users teleoperating a pair of 7-degree-of-freedom robotic arms through a pick-and-place task in VR via two control systems: a stationary 3D mouse and via the VR controllers that tracked participants’ movements. Faster execution times were gained when using the VR controller rather than the 3D mouse. Similarly, Martin-Barrio and colleagues (2020) compared different control systems for teleoperating a robot in VR, including controller, Master-Slave, and physical gestures. Particularly, users maneuvered the robotic arm Kyma with high degrees of freedom to some predetermined positions. Physical gestures were preferred over Master-Slave and controller, and they additionally allowed higher accuracy and faster operation times compared to the controller. On the other hand, in the same study, participants self-reported lower values of workload when using a controller and direct manipulation compared to the master-slave modality.

Besides, there are also studies that did not show an advantage of HICSs over LICSs. For instance, in Rouanet et al. (2009), participants had to teleoperate a zoomorphic robot through a domestic environment to find an object. They were instructed to use three control systems: touchscreen-based buttons, a virtual keyboard on a 2D screen, and arm movements tracked by a hand-held controller. The authors did not observe any performance differences between the input modalities; however, participants preferred to use the touchscreen-based input modality over the other two. In the experiment of Grabowski and colleagues (2021), instead, participants had to drive a mobile robot equipped with two arms through a pick-and-place task in an immersive large-scale virtual environment. In one condition, they teleoperated the robot via VR controller buttons, while in the other condition they had to physically walk to the target position. In both cases, participants teleoperated the robotic arms by moving their own arms. Results demonstrated a better performance in task completion time and accuracy when using the VR controllers rather than when active walking. Overall, studies assessing users’ performance when teleoperating robots via control systems allowing higher or lower interactivity show a general tendency for better performance when using HICSs, even though there is some evidence in contrast with this trend (Rouanet et al., 2009; Grabowski and colleagues, 2021). Differently, literature on the effects of the degree of interactivity on users’ workload is still scarce.

## 2.2 Eye-tracking evaluations in robotics and teleoperations

Eye-tracking metrics were often used to predict workload in human-machine interactions (e.g. during driving or in air traffic control operations, McIntire et al., 2014; Ahlstrom and Friedman-Berg, 2006), and in fewer cases also in the robotics domain (e.g., Novak et al., 2015; Nenna et al., 2022). Several studies demonstrated how pupil diameter increases with increasing cognitive workload (Kahneman, 1973; Marinescu et al., 2018; Pomplun and Sunkara, 2019). Pupil diameter variations are related to the activity of the parasympathetic nervous system, which is involved in the regulation of arousal (Köles, 2017). However, light has a significant influence on pupil size variations as well, making it crucial to keep constant lighting of the setting, pre-processing pupil data, and apply proper baseline correction in order to exclude pupil size variations that are possibly unrelated to the user’s cognitive activity (Mathot et al., 2018; Nenna et al., 2022). Evidence of the relation between pupil diameter and workload in robotics was repeatedly found in the surgical domain, where participants were asked either to perform a robotic surgical task under different difficulty levels (Wu et al., 2020) or to bring rubber objects over dishes with different target sizes and distances (Zheng et al., 2015). In the industrial field, instead, Nenna et al. previously demonstrated how pupil diameter variates with task load in participants physically executing a pick-and-place task with a robotic arm (Nenna et al., 2022). This was true both when operating on the physical robot (Universal Robot e-Series UR5e) and on its digital counterpart in VR. Additionally, smaller pupil size variations were also observed when interacting with the virtual compared to the physical robot, suggesting how the virtual robot was preferable to the physical one as it allowed to save mental resources.

Furthermore, PERCLOS is a robust measure of vigilance for humans interacting with machines, particularly in the automotive area (e.g., Du et al., 2022). PERCLOS can be defined as the percentage of time that the eyelids cover the eye area by more than 80% and can be thus gained from continuous data on eye openness. Literature on this metric showed that higher levels of fatigue and lower vigilance are associated with a higher PERCLOS (for a review, Marquart et al., 2015). However, to the best of our knowledge, there is no study demonstrating the reliability of PERCLOS as a measure of vigilance or fatigue in the robotic context. Indeed, in the previously cited study of Wu et al. (2020), only pupil diameter and gaze entropy differed between different task difficulty levels, while PERCLOS did not show any significant variation between the conditions.

Blinks can also be informative of one's level of workload (Fogarty and Stern, 1989; Marquart et al., 2015) and/or fatigue (Kim et al., 2022). For example, blink frequency was demonstrated to be inversely related to the level of mental load (Holland and Tarlow, 1972; Zheng et al., 2012; Borghini et al. 2014), and to the performance in a static simulated air traffic vigilance task (McIntire et al., 2014). Similarly, during air traffic control operations, a decrease in blink duration was observed in case of increased visual workload (Ahlstrom and Friedman-Berg, 2006), while increased blink duration was associated with a deterioration of performance in a vigilance task (McIntire et al., 2014). A possible explanation is that, under high mental demand, users tend to inhibit blinks to reduce the risk of missing incoming information (Fogarty and Stern, 1989). Evidence in this direction was also found when executing a simulated laparoscopic task: fewer and shorter blinks were associated with a higher workload as self-reported at the NASA-TLX score (Zheng et al., 2012). More recently, Guo et al., 2021 evaluated the mental workload during a space robot teleoperation: participants controlled a robotic arm via desktop and joystick under varied latency and time pressure, which is known to affect workload. When time pressure increased, blink frequency decreased and pupil size increased, while no substantial differences were observed across latency manipulations.

Some studies also showed how eye-tracking measures could predict workload during HRIs. For instance, Novak et al. (2015) investigated the sensibility of different continuous eye-tracking indexes in estimating workload via machine learning, proposing a continuous inference rather than a classification into discrete classes of workload indexes. During the task, targets containing equations were presented on a screen for a few seconds. Using the ARMin robot, participants had to hit only the targets that contained the correct equations. In this context, eye-tracker metrics were able to identify progressive increases in workloads induced by a gradual increase in the task's difficulty. Specifically, pupil dilatation was the most sensitive index to workload, while blink and fixation frequencies were the most sensitive to effort. Similar results were obtained by Gao et al. (2013), who compared the ability of different workload-related eye measurements (including blink parameters and pupil dilatation) in predicting the overall mental workload as self-reported via the NASA-TLX questionnaire during digital nuclear power plant operations. Interestingly, none of the single measures was reliable in assessing the overall mental workload. However, when integrating all the measures within a predictive model, the overall mental workload was assessed accurately. Furthermore, blink rate demonstrated higher sensitivity to workload, while pupil size was more sensitive to error-related attention and arousal.

# 3. METHODS

## 3.1 Sample

As suggested by the power analysis conducted on Gpower (Erdfelder et al., 1996), for our within-subjects experimental design, a total sample of 21 participants was needed to detect a medium effect size (d = 0.5) with 80% power. 24 participants, 11 females and 13 males ($M_{age}$ = 26.16; $SD_{age}$ = 1.85), voluntarily took part in the experiment after signing informed consent. They all reported having normal or corrected-to-normal visual acuity (via contact lenses), normal color vision, and no current or past neurological or psychiatric problems. The experimental protocol was approved by the local ethics committee, and the study was conducted following the principles of the Declaration of Helsinki. Two participants were excluded for technical issues of the eye tracker, while one more participant withdrew from the experiment because was reporting visual difficulties in VR. Finally, three participants were excluded from the analysis for having committed more than 50% of errors in the arithmetic task. The final sample comprised 18 participants, 9 females and 9 males ($M_{age}$ = 26.33; $SD_{age}$ = 2.02).

## 3.2 Technical set-up

An HTC Vive Pro Eye headset (resolution: 1440 x 1600 pixels per eye; refresh rate: 90 Hz; Field of view: 110°) was connected to an MSI laptop (model GT63 Titan 8RF, processor Intel Core i7-6700HQ, RAM 16Gb). This model of head-mounted display has an eye-tracking system embedded (sampling frequency: 120 Hz; calibration: 5-points) which allows continuous recording of eye parameters. Based on the present research questions, the virtual environment (Fig. 1) has been re-arranged from the one previously developed by Nenna et al. (2022), which was programmed in Unity (version 2019.4.18f1). At the end of each experimental session, a large datalog was automatically saved on the MSI laptop, which included time series data of position and rotation of the VR headset and controllers, of all interactions between the user and the robot, and of the accuracy of the performed pick-and-place task.

## 3.3 Task and procedure

All participants signed informed consent before starting the experiment. Thereafter, they filled out questionnaires about their demographics, VR expertise and general preferences for virtual robot control systems. Specifically, we asked: “If you had to guide a robotic arm in VR, which of the following control modalities would you prefer?”. The possible answers were “Controller buttons” and “Physical actions”. Afterward, all participants underwent a training session to familiarize with the tasks, in which they performed a few trials of each of them. All instructions were presented in text format in the virtual environment. Once the participant reported having understood all the tasks, a 5-point calibration of the eye-tracking system was conducted, and the experiment started.

During the experiment, participants performed 5 tasks composed of 40 trials each that were re-adapted from previous research (Nenna et al., 2022): (1) an arithmetic task, (2) a pick-and-place task executed both via controller buttons (button-based control systems), (3) and physical actions (action-based control system), (4) and a dual-task performed via controller buttons (button-based control systems), (5) and physical actions (action-based control system) as well. These tasks were presented in a random fashion, and a NASA-TLX questionnaire was administered at the end of each task. Participants could also claim a break after each NASA-TLX questionnaire; in that case, the eye-tracking system was re-calibrated before starting the next task. Once participants finalized all the tasks, the final question on the general preferences for virtual robot control systems was administered again ("With reference to the experience you have just concluded, which of the following control modalities did you prefer?") and the experiment ended.

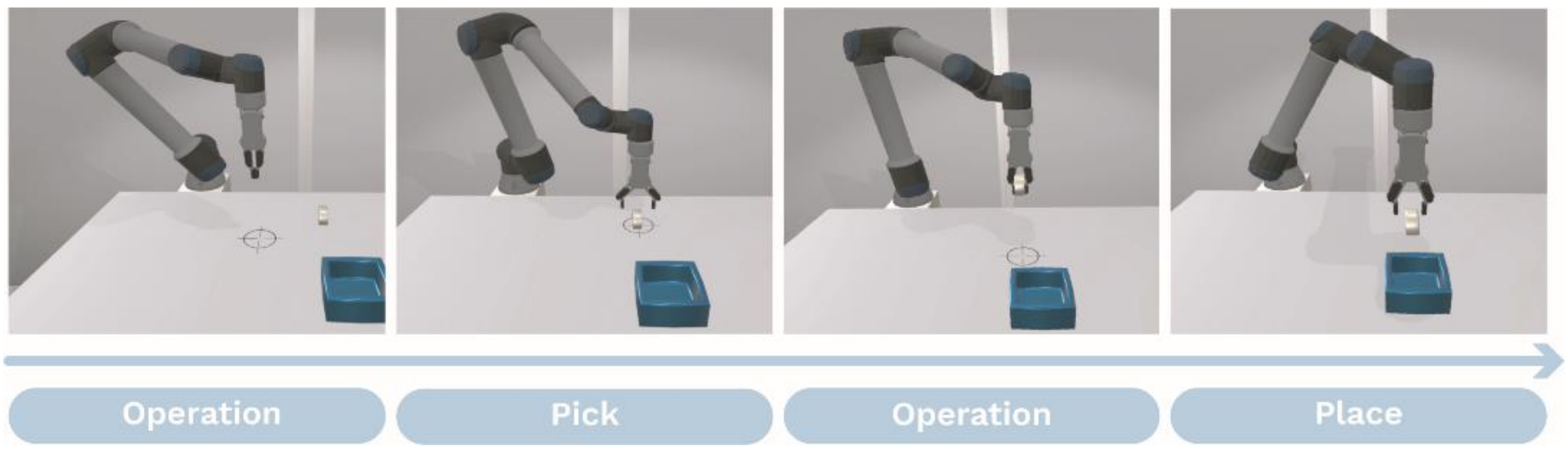

**Figure 1**. Overview of the pick-and-place task. In the Operation phases, the participant is driving the robot toward the bolt or the box. In the Pick phase, the participant executed the command for picking the bolt, and in the Place phase, for placing the bolt inside the box.

In the arithmetic task (1), participants mentally summed a series of four numbers presented in text format in the virtual environment. The numbers appeared on a virtual panel that always followed the participant's head movements and was placed in the upper part of his/her view, in a way that it could not cover the work table nor the robot's effector and was always inside the participant's functional field of view. Thereafter, they reported the result of the arithmetic operation in a virtual keyboard by using the controller buttons. The presented numbers randomly ranged between 1 and 10 and a time interval of 2.5sec ± 0.3sec intercurred between them. For the pick-and-place task, we used the same paradigm employed by Nenna et al., (2022), in which participants were asked to guide the robotic arm to pick a bolt from the workstation and place it into a box (Figure 1). While the task design remained unchanged, we here introduced two different control systems: in the button-based pick-and-place task (2), participants used the pad button of the right controller to move the robot on the left-right-farward-backward directions over the work table; in the action-based pick-and-place task (3), instead, they were allowed to reach the virtual robot with their right hand, grasp it by pressing the grip button on the right controller and then move it to the desired position by simply moving their own arm. The latter condition thus reproduced the direct manipulation feature of cobots. In both conditions, once the robot was placed in the right location, participants pressed the pad button on the left controller for picking or placing the bolt. Therefore, participants used the VR controllers with both control systems, but only in the action-based condition they were allowed to interact with the virtual robot physically. Finally, in the dual-task, the pick-and-place task and the arithmetic task were concurrently performed, once using the button-based (4) and once using the action-based (5) control systems. The series of numbers presented for the arithmetic task covered the whole pick-and-place task duration, and the result was reported only after the bolt was placed into the box. All task conditions are depicted in Figure 2.

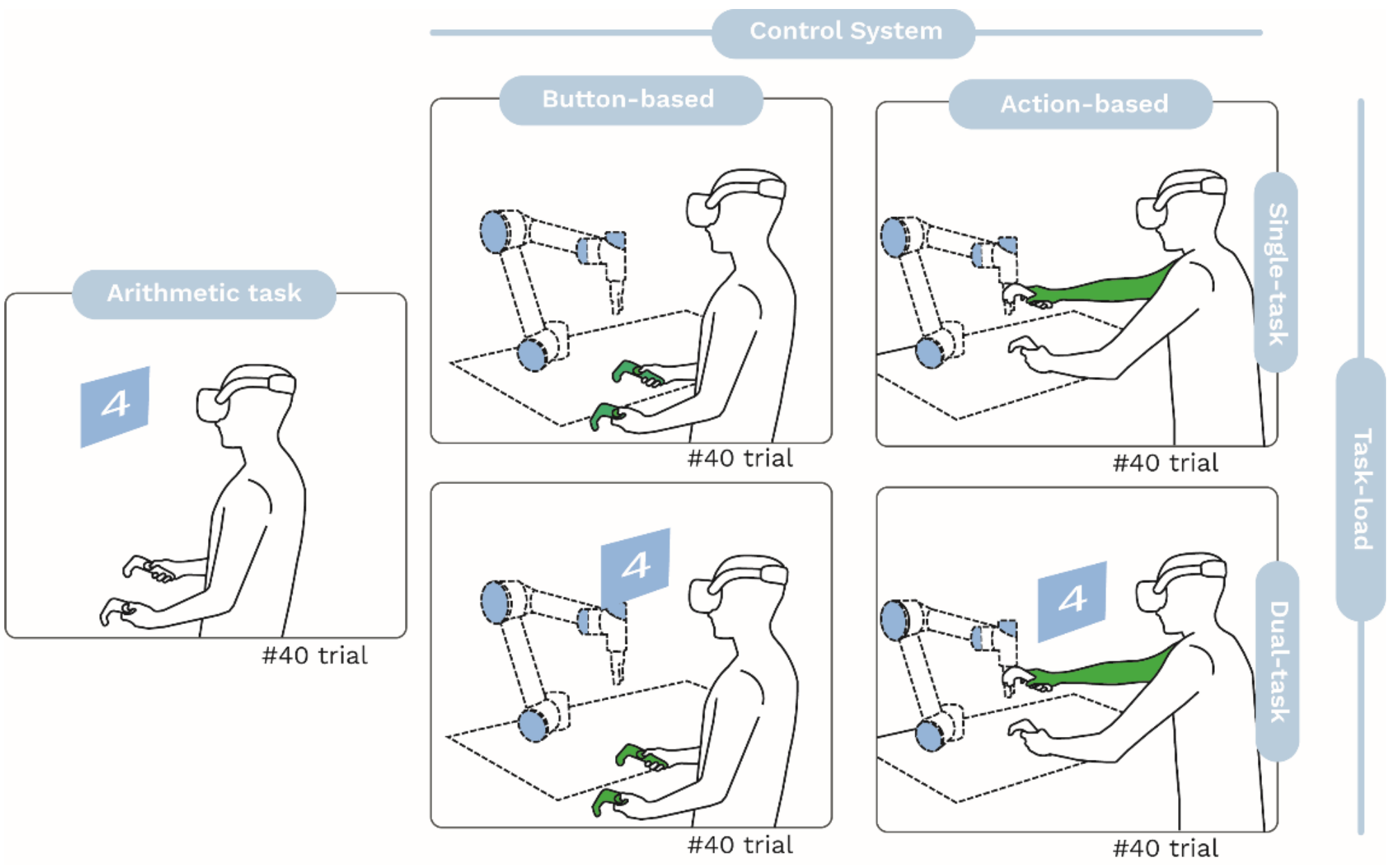

**Figure 2**. Experimental design. All participants executed an arithmetic task as a baseline, and a pick-and-place task under low (single-task) and high (dual-task) mental load. The pick-and-place was additionally executed under two control system conditions: low-interactivity and high-interactivity. All participants executed all five tasks.

## 3.4 Measurements

### 3.4.1 Pick-and-place performance

We measured the operation times as the time elapsed during the robot's movements (start: first movement of the robot; end: last movement before the pick/place action). Considering that the pick action required higher precision for aligning the robot effector with the bolt to pick as compared to the place action, users' performance was analyzed independently in the pick and in place phases. Trials whose duration exceeded 4 SD from the averaged duration were removed as they represented very unrealistic operation times (pick phase: 1.43% removed, range 13.45sec-46.99sec; place phase: 0.08% removed over 2.827, range 14.91sec-42.94sec). The same trials were not considered for the analysis of the other independent measures either. Additionally, the error rate at the pick-and-place was measured independently for the pick and the place phases. Particularly, the pick and place automations were executed only if the left pad button was pressed while the robot was perfectly positioned above the bolt in the pick phase, and above the box in the place phase ("correct" event). If at the first attempt of button press the robot was not in line with the bolt/box, the event was registered as "incorrect". The participant thus had to relocate the robotic arm in the right position to initiate the automation. The percentage of "incorrect" events registered for each action informed on the error rate at the pick and at the place actions.

### 3.4.2 Arithmetic performance

In the arithmetic task, we measured the arithmetic input time as the time elapsed from the end of the last number presentation to the moment the participant sent the result of his/her arithmetic calculation through the controller. This measure can be informative of the mental effort deployed for finalizing the mental calculations in each condition. When computing it, we only considered trials in which the correct sum was sent. The reason behind this choice is that in case participants lost count of the sum during the arithmetic task, they often quickly insert a casual number, thus resulting in casual input time. Differently, trials in which participants sent the correct result are more likely to be the product of meaningful cognitive processes. Finally, the error rate at the arithmetic task was computed as the percentage of wrong sums reported, allowing an understanding of the dual-task-induced interference with the main task (pick-and-place task).

#### 3.4.3 Eye-tracking measures

Based on previous studies (Guo et al., 2021; Nenna et al., 2022; Novak et al., 2015; Wu et al., 2020), we computed pupil size variation, perclos, blink frequency and duration as workload-related eye-tracking indexes. Specifically, pupil data preprocessing followed the same procedure used in Nenna et al. (2022). The pupil size values were averaged across the left and right eye and a median filter was applied. We ensured that none of the trials nor the participants had more than 35% missing data within the analyzed time windows and then applied a subtractive baseline correction on the first 4 data points at trial level (corresponding to about 200ms on average). Therefore, only pupil size variations compared to the baseline period were assessed (Mathôt et al., 2018b). For the pick-and-place task, variations in pupil size were analyzed within the pick and the place actions independently, whose length was standardized via dynamic time warping to fit 30 data points. For the arithmetic task, instead, we selected four time windows, one for each number presented, and applied dynamic time warping to standardize their length (which ranged between 2.3 and 2.7sec). Additionally, we used the eye openness data stream outputted from the HTC Vive headset for computing perclos and blinks. perclos was calculated as the percentage of time during which the eyelids covered pupils by more than 80% of their area (Wu et al., 2020) in four time windows, each including 10 trials. Blinks were detected as eye closures lasting a minimum of 70ms and a maximum of 500ms (Benedetto et al., 2011). If eyes were closed for less than 70ms, it was considered a technical issue of the eye-tracker that likely lost pupil tracking for some frames (Faure et al., 2016). Blink frequency was operationally defined as the blink rate per minute.

#### 3.4.4 Self-reports

The NASA-TLX questionnaire was administered after each task as a measure of self-reported workload. Once before starting the experiment and once in the end, we additionally administered a question asking to express the individual preference for guiding a robot in VR either via controller buttons or physical actions. With these questions, we intended to measure whether the individual preferences for one or the other control system would change after having tested both the button-based and the action-based control systems.

### 3.5 Statistical Analysis

#### 3.5.1 Performance measures

For analyzing performance data, we used Generalized Linear Models (GLMs from *lme4* package, Bates et al., 2014) in RStudio (Team, 2022). Data were first fitted through the function *descdist()* of the package *fitdistrplus* (Delignette-Muller and Dutang, 2015), then the appropriate models were chosen accordingly to data distribution. For each of the performance measures at the pick-and-place task, we computed a GLM including the factors Task load (single-task, dual-task) and Control System (button-based, action-based) with Participant as a random effect. For analyzing performance measures at the arithmetic task, instead, we run a GLM over the factor Task (single-task, button-based dual-task, action-based dual-task). The Bonferroni correction was always applied when interpreting the post hoc contrasts within the significant interactions.

#### 3.5.2 Eye-tracking measures

GLMs from *lme4* package (Bates et al., 2014) were also used for the analysis of eye-tracking data in RStudio (Team, 2022). Specifically, each model was chosen based on data distribution (Delignette-Muller and Dutang, 2015). Models analyzing pupil size variation during the pick-and-place included the factors Task Load (single-task, dual-task), Control System (button-based, action-based) and Window (1, 2, 3, 4, 5, 6) with Participant as a random effect. The factor Window allowed to consider pupil size variation changes in the time course on trial level. When analyzing the pupil size variation throughout the arithmetic task, instead, we ran a model including the factors Task (arithmetic task, button-based dual-task, action-based dual-task) and Arithmetic operation (start, 1st sum, 2nd sum, 3rd sum), with Participant as a random effect. For this analysis, we only considered the first 3 arithmetic operations in order to compare the Single-task with the Dual-tasks. Differently, the statistical models analyzing perclos, blink frequency and duration included the factors Task Load (single-task, dual-task), Control System (button-based, action-based) and Window (1, 2, 3, 4) with Participant as a random effect. Each window included 10 trials (window 1: trials 1-10; window 2: trials 11-20, etc.) and allowed to look into eye parameters' changes in the time course on task level.

#### 3.5.3 Self-reports

The analysis of the NASA-TLX questionnaire was conducted through a GLM that included the following factors: Task Load (single-task, dual-task), Control system (button-based, action-based), and Items (mental demand, physical demand, temporal demand, performance, effort, frustration), with Participant as a random effect. Post hoc contrasts were performed specifically between the levels of Task load and Control system in each of the questionnaire's items, with the application of the Bonferroni correction (Bonferroni et al., 1936). Relations between the NASA-TLX score (both the overall score and the score at the individual NASA-TLX items) and each of the eye-tracking measures were also assessed via Pearson's linear correlation tests. Furthermore, we reported the response rate of the individual preference for action- vs. button-based control systems expressed before and after the experiment.

# 4. RESULTS

## 4.1 Performance measures

#### 4.1.1 Pick-and-place performance

The GLM conducted on the operation times demonstrated significant main effects of Task Load only in the pick phase ($X^2$ = 20.02, p<.0001), but not in the place phase ($X^2$ = 20.02, p=.053), while a main effect of Control System was observed both in the pick ($X^2$ = 1462, p<.0001) and in the place ($X^2$ = 1976.7, p<.0001) phases. Additionally, interaction effects between Task Load and Control System were observed only for the place ($X^2$ = 10.64, p<.01) but not for the pick action ($X^2$ = 0.56, p=.45). Post-hoc contrasts revealed significant differences between the button- and action-based control systems both under single- (p<.0001) and dual-task (p<.0001). After applying the Bonferroni correction, differences between the single- and dual-task were not significant in any of the control system modalities. When analyzing the pick-and-place error rate, results demonstrated significant main effects of Task Load ($X^2$ = 5.91, p<.05) and Control System ($X^2$ = 22.27, p<.0001) only in the pick but not in the place phase. No interaction effects were observed. All performance results are depicted in Figure 3 and summarized in Table 1.

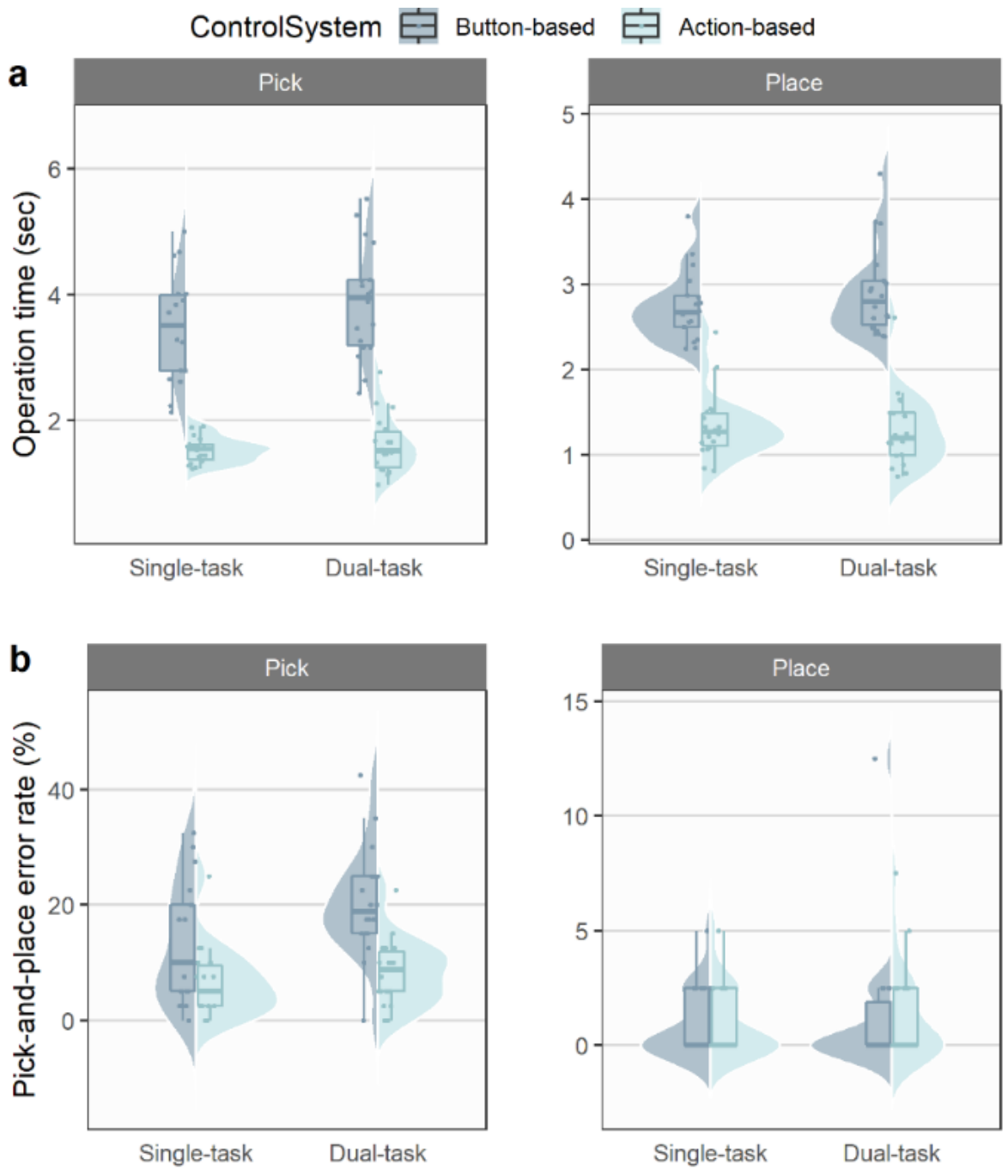

**Figure 3**. Averaged operation time (a) and error rate (b) at the pick-and-place task. In each plot and condition, a boxplot and a half violin depict the data distribution. Each point corresponds to the averaged data of one participant.

**Table 1. Descriptive statistics of the performance at the pick-and-place task**

| | | Operation time (sec) | | Error rate (%) | |
|---|---|---|---|---|---|
| | | Pick<br>Mean (SD) | Place<br>Mean (SD) | Pick<br>Mean (SD) | Place<br>Mean (SD) |
| Task Load | Single-task | 2.51 (1.88) | 2.04 (1.18) | 9.93 (9.03) | 0.97 (1.50) |
| | Dual-task | 2.73 (2.00) | 2.11 (1.28) | 14.2 (9.91) | 1.32 (2.57) |
| Control System | Button-based | 3.67 (2.07) | 2.83 (1.07) | 16.8 (10.4) | 1.11 (2.35) |
| | Action-based | 1.57 (1.57) | 1.31 (0.85) | 7.36 (5.91) | 1.18 (1.84) |

### 4.1.2 Arithmetic performance

The GLM on the arithmetic error rate indicated a significant main effect of the factor Task ($X^2$ = 14.58, p <.001). Post-hoc contrasts revealed that, compared to the single arithmetic task, the error rate was significantly higher only while executing the pick-and-place task via Low-Interactivity ($p<.01$) but not via High-Interactivity control system ($p=.16$). Moreover, the error rate at the arithmetic task did not differ significantly between the two dual-tasks ($p=.39$). For the analysis of the arithmetic input time, instead, the main factor Task resulted in being statistically significant ($X^2 = 146.04$, $p<.0001$). Post hoc contrasts revealed that the arithmetic input time was significantly lower in the Single-task compared to both the Dual-task conditions ($p_s<.0001$). Moreover, a significantly higher arithmetic input time was observed in the button-based dual-task condition compared to the action-based dual-task condition ($p<.0001$). Performance results at the arithmetic task are depicted in Figure 4 and summarized in Table 2.

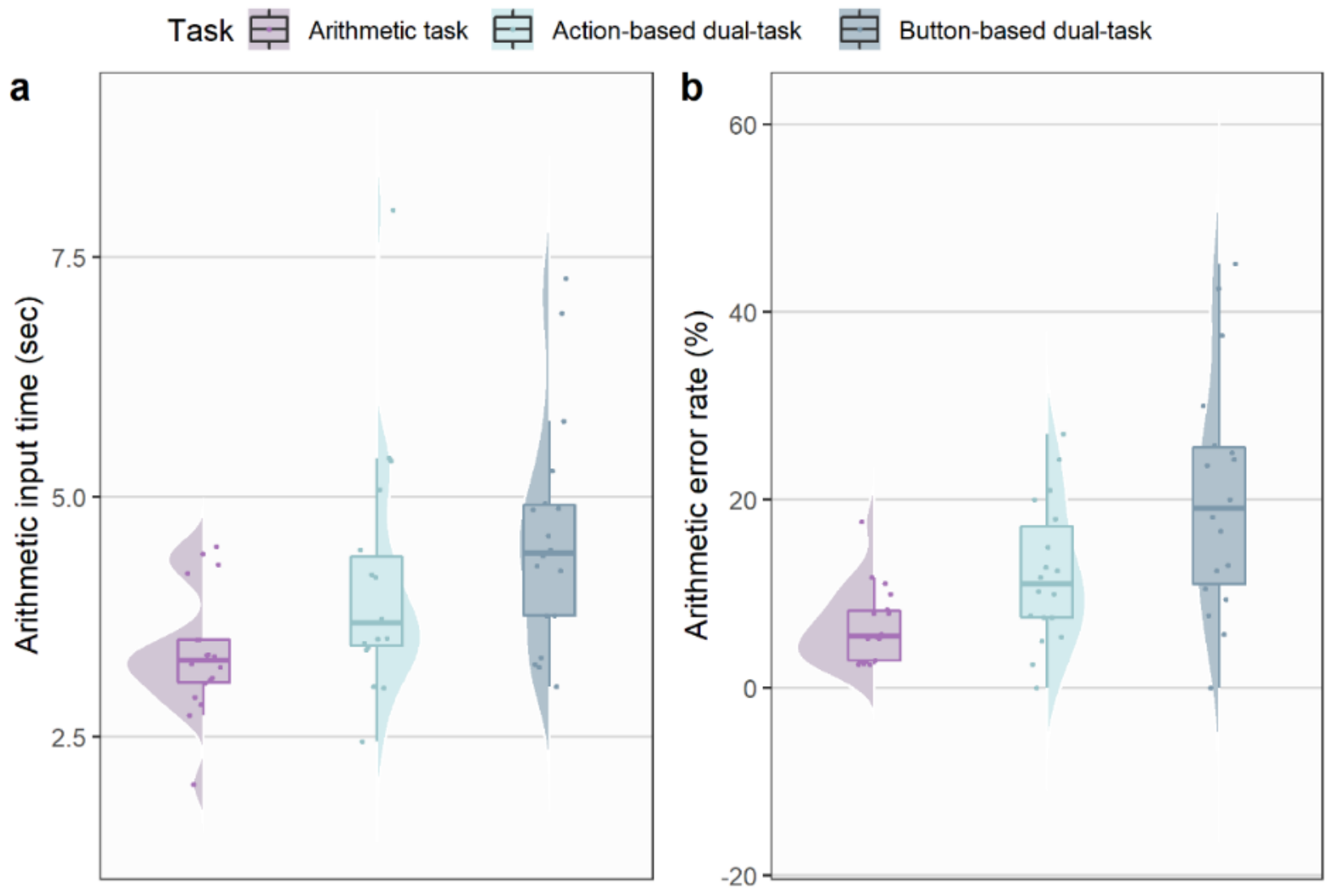

**Figure 4**. Averaged input time (a) and error rate (b) at the arithmetic task. In each plot and for each task condition, a boxplot and a half violin depict the data distribution. Each dot corresponds to the averaged data of one participant. Abbreviation: DT = Dual-task.

| **Table 2. Descriptive statistics of the performance at the arithmetic task** | | | |
|---|---|---|---|
| | | Error rate (%) | Input time (sec) |
| Task | Arithmetic single-task | 6.55 (4.10) | 3.37 (1.49) |
| | Action-based dual-task | 12.1 (7.52) | 3.95 (2.16) |
| | Button-based dual-task | 20.4 (12.6) | 4.38 (2.28) |

## 4.2 Eye-tracking measures

### 4.2.1 Pupil size variation - Arithmetic task

As depicted in Figure 5, the analysis of pupil size at the arithmetic task yielded significant main effects of both Task ($X^2 = 1553.6$, $p <.0001$) and Arithmetic operation ($X^2 = 1017.3$, $p <.0001$). Furthermore, two factors interacted significantly ($X^2 = 1346.3$, $p <.0001$). Post-hoc contrasts Bonferroni-corrected showed a significant increase in pupil size when moving from Start to 2nd sum only for the button-based dual-task ($p < .001$) and from Start to 3rd sum for all task conditions (all $p_s$ <.001). Similarly, significant pupil size increases were observed when moving from the 1st sum to 2nd for the arithmetic task (p<.01) and the button-based dual-task (p<0001), and from 1st sum to 3rd sum for all tasks (all $p_s$ <.0001). When moving from 2nd to 3rd sum, only the dual-task conditions yielded significant contrasts (all $p_s$ <.0001).

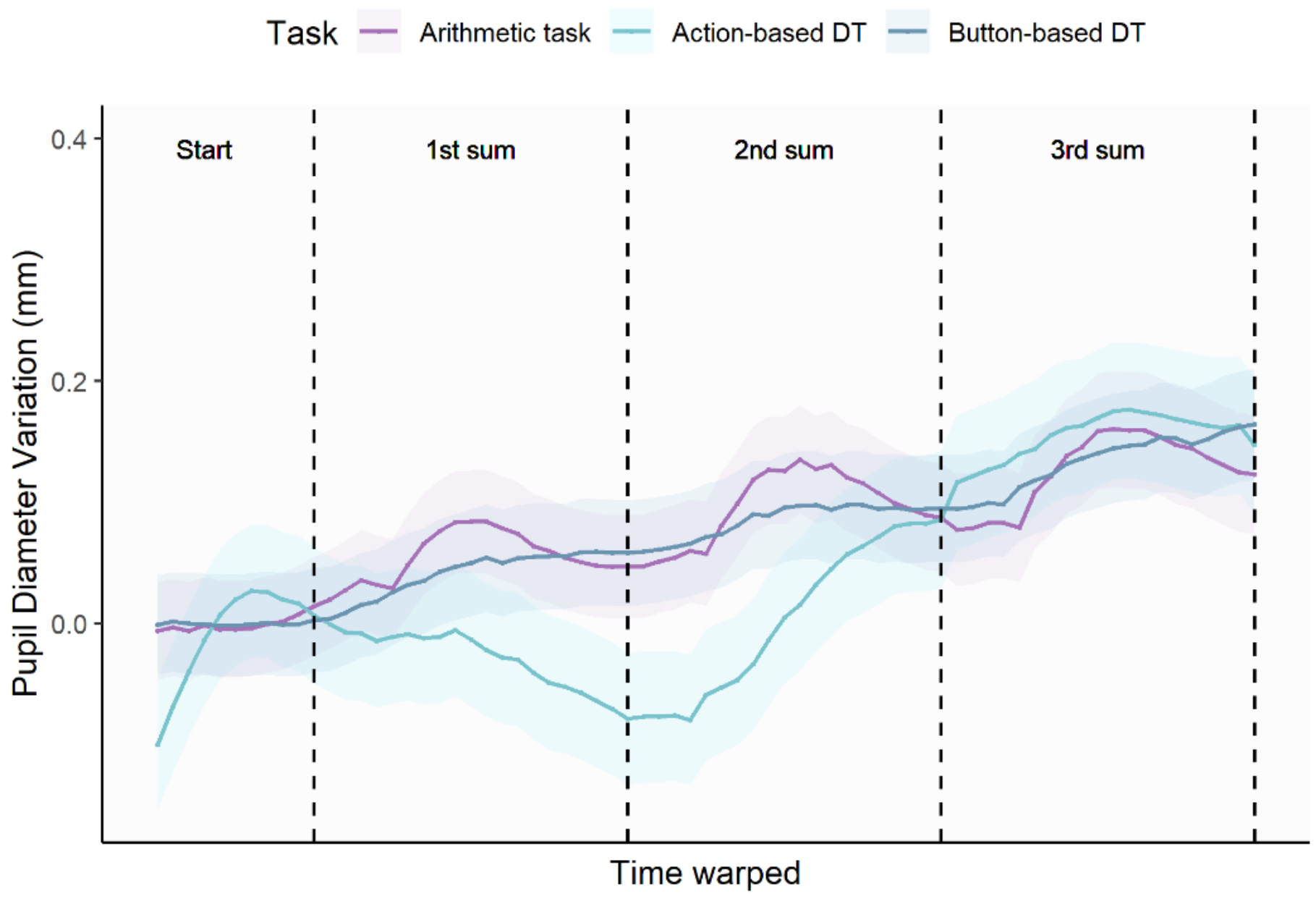

**Figure 5**: pupil size variations throughout the arithmetic task, grouped by Task (Arithmetic task, Action-based dual-task, Button-based dual-task). Lines represent the averaged pupil size in each data point, and the shadows indicate the standard error. Abbreviation: DT = Dual-task.

### 4.2.2 Pupil size variation - pick-and-place task

The statistical model yielded significant interactions between Window and both Task Load (pick: $X^2$ = 981.76, p<.0001; place: $X^2$ = 382.18, p<.0001) and Control System (pick: $X^2$ = 1884.75, p<.0001; place: $X^2$ = 819.96, p<.0001), and between Window, Task Load and Control System (pick: $X^2$ = 52.20, p<.0001; place: $X^2$ = 66.23, p<.0001). Post hoc of interest included the comparison between Single- and Dual-task in each Control System condition and within each Window. Specifically, pupil size variation was significantly higher in the dual-task compared to the single-task in both Control System conditions and from window 2 to 6 specifically in the pick phase (all $p_s$ < .0001) and in the button-based condition of the place phase (all $p_s$ < .0001). Differently, in the action-based condition of the place phase, pupil size variation was higher in the dual-task compared to the single-task in windows 4 (p<.01), 5 and 6 ($p_s$ < .0001). Results of pupil size variation are depicted in Figure 6.

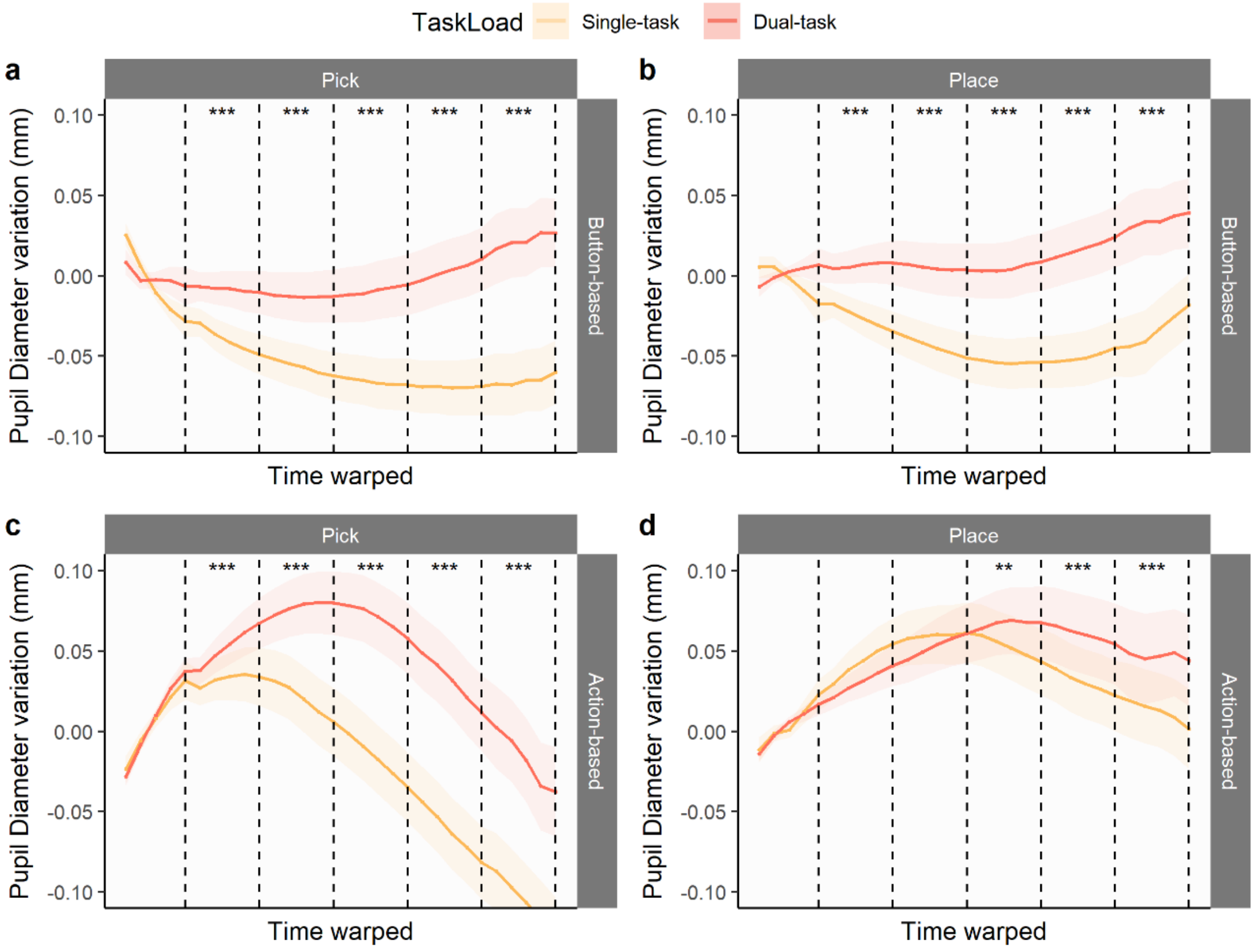

**Figure 6**. Pupil size variations during the pick-and-place task relative to the Task load conditions. The first row depicts the pick (a) and place (b) phases of the button-based condition. The second row depicts the pick (c) and place (d) phases of the action-based condition. All the plots are complemented by stars indicating the significance level of the post hoc tests (*p ≤ .05; **p ≤ .01; ***p ≤ .0001)

### 4.2.3 Perclos

A main effect was observed only for Task Load ($X^2$ = 16.55, p<.0001). Interactions with the factor Window did not reach the significance threshold not for Task Load ($X^2$ = 0.27, p=.96) nor for Control System ($X^2$ = 2.06, p=.55). Only the interaction between Task Load and Control System was statistically significant ($X^2$ = 4.71, p<.05), with post hoc contrasts revealing significant differences between single- and dual-task only in the action-based condition (p<.0001). Results on the perclos are depicted in Figure 7.

### 4.2.4 Blink parameters

The analysis of blink duration yielded a significant main effect only for the factor Window ($X^2$ = 14.2, p<.01), while Task load ($X^2$ = 1.29, p=.25) and Control system ($X^2$ = 3.05, p=.08) were not significant. A significant interaction effect was observed only between Task load and Control System ($X^2$ = 6.69, p<.01); however, after applying the Bonferroni correction, none of the contrasts reached the significance threshold. Differently, when analyzing blink frequency, the factors Task load ($X^2$ = 12.32, p<.001) and Window ($X^2$ = 31.79, p<.0001) were demonstrated to be statistically significant. Specifically, higher blink frequency was observed in the single task (M = 3.99, SD = 3.7) compared to the dual-task (M = 2.41, SD = 2.43). Furthermore, a significant interaction effect was observed between Task load and Control system ($X^2$ = 10.52, p<.01), with higher blink frequency observed in the single- compared to the dual-task only in the action-based (p<.001) but not in the button-based condition (p=0.54). Results on blink parameters are shown in Figure 7.

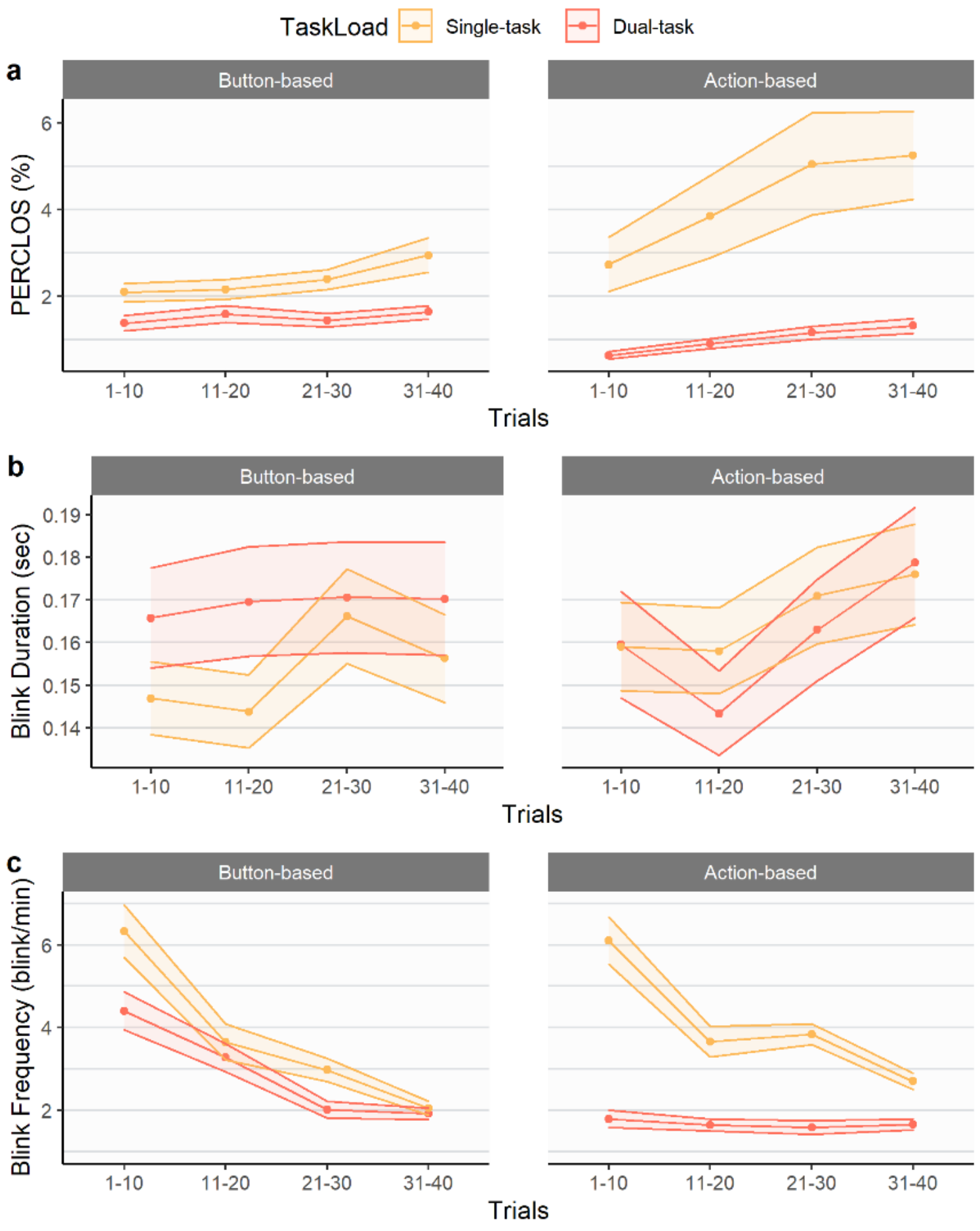

**Figure 7**: Perclos, blink duration and frequency throughout the experimental sessions in each Task Load (single-task, dual-task) and Control system condition (button-based, action-based).

## 4.3 Self-reports

### 4.3.1 NASA-TLX questionnaire

Results on the NASA-TLX questionnaire, as depicted in Figure 8, yielded a significant main effect of both Task Load ($X^2$ = 212.99, p<.0001) and Control system ($X^2$ = 12.79, p<.001). Significant interaction effects were also observed between Item and Task Load ($X^2$ = 47.32, p<.0001) and Item and Control system ($X^2$ = 11.10, p<.0001). Post hoc contrasts revealed significant differences between Single- and Dual-task for the following items: Mental Demand (p<.0001), Temporal Demand (p<.0001), Performance (p<.05), Effort (p<.0001), and Frustration (p<.0001). Differently, significant differences between the button-based and action-based conditions were only observed for the item Frustration (p<.01).

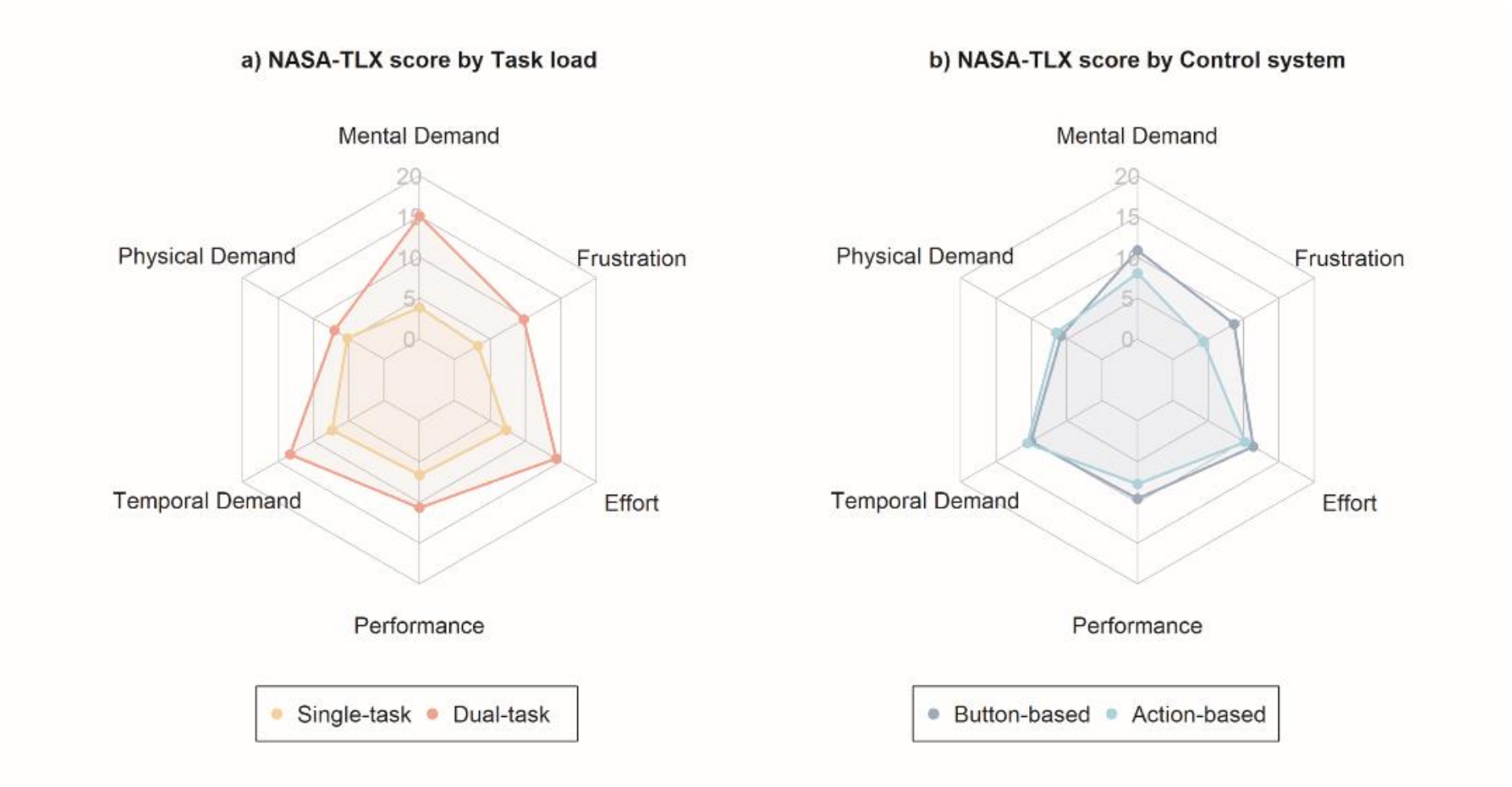

**Figure 8**. Averaged NASA-TLX score in each item according to the Task load (a) and Control system (b)

### 4.3.2 Individual preferences for button- vs. action-based control systems

Finally, individual preferences for button- or action-based control systems expressed before and after the experiment are shown in Figure 9. Before the experiment, 73.68% of participants expressed a preference for guiding the robot via physical actions, and after the experiment, their percentage increased to 89.47%.

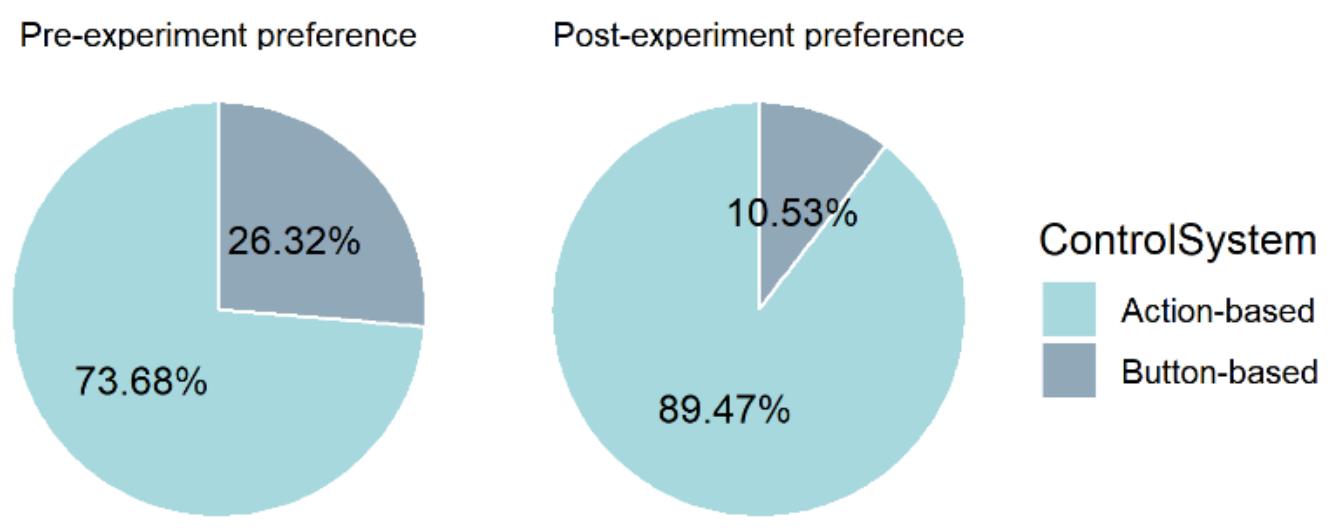

**Figure 9**. Pre- and post-experiment preferences for operating through action- and button-based Control systems expressed by participants before and after the experimental session

## 4.4 Relations between self-reported workload and eye-tracking indexes

Results of the correlation matrix are reported in Figure 10. Independently from the level of interactivity, we observed that blink duration had only a significant inverse relation with the self-reported mental demand ($R = -0.28$, $p = .026$). Similarly, perclos showed an inverse correlation with the overall NASA-TLX score ($R = -0.31$, $p = .013$), and also with mental demand ($R = -0.25$, $p = .042$) and effort ($R = -0.31$, $p = .011$). Pupil size additionally demonstrated a positive correlation with the overall score at the NASA-TLX ($R = 0.28$, $p = .022$), and the single following dimensions: mental demand ($R = -0.36$, $p = .003$), temporal demand ($R = 0.29$, $p = .021$) and effort ($R = 0.31$, $p = .011$). It is to notice that all the latter were weak relations.

Differently, when differentiating for the two control systems, we observed stronger relations between self-reported workload and eye parameters. Specifically, in the button-based condition, pupil size was strongly correlated both with the overall workload ($R = 0.6$, $p < .001$) and the self-reported mental demand (R = 0.66, p < .0001). Furthermore, moderate positive relations were observed between pupil size and effort (R = 0.58, p < .001) and frustration (R = 0.46, p < .01). Differently, perclos demonstrated only weak negative relations with performance (R = -0.39, p = .027) and temporal demand (R = -0.39, p = .029), while blink frequency was moderately correlated with performance (R = -0.44, p = .012). Blink duration did not show any significant correlation with any of the self-reported workload dimensions.

When looking at the same relations in the action-based condition, we notice significant strong relation between pupil size and mental demand (R = 0.66, p < .0001), and moderate relations between pupil size and frustration (R = 0.55, p = .0016), effort (R = 0.41, p = .018) and even the overall NASA-TLX score (R = 0.43, p = .014). Perclos was negatively related to the overall workload (R = -0.38, p = .031) and with the self-reported effort (R = -0.39, p = .026). Finally, blink frequency and duration demonstrated moderate negative relations with the mental demand (blink frequency: R = -0.48, p = .005; blink duration: R = -0.40, p = .023).

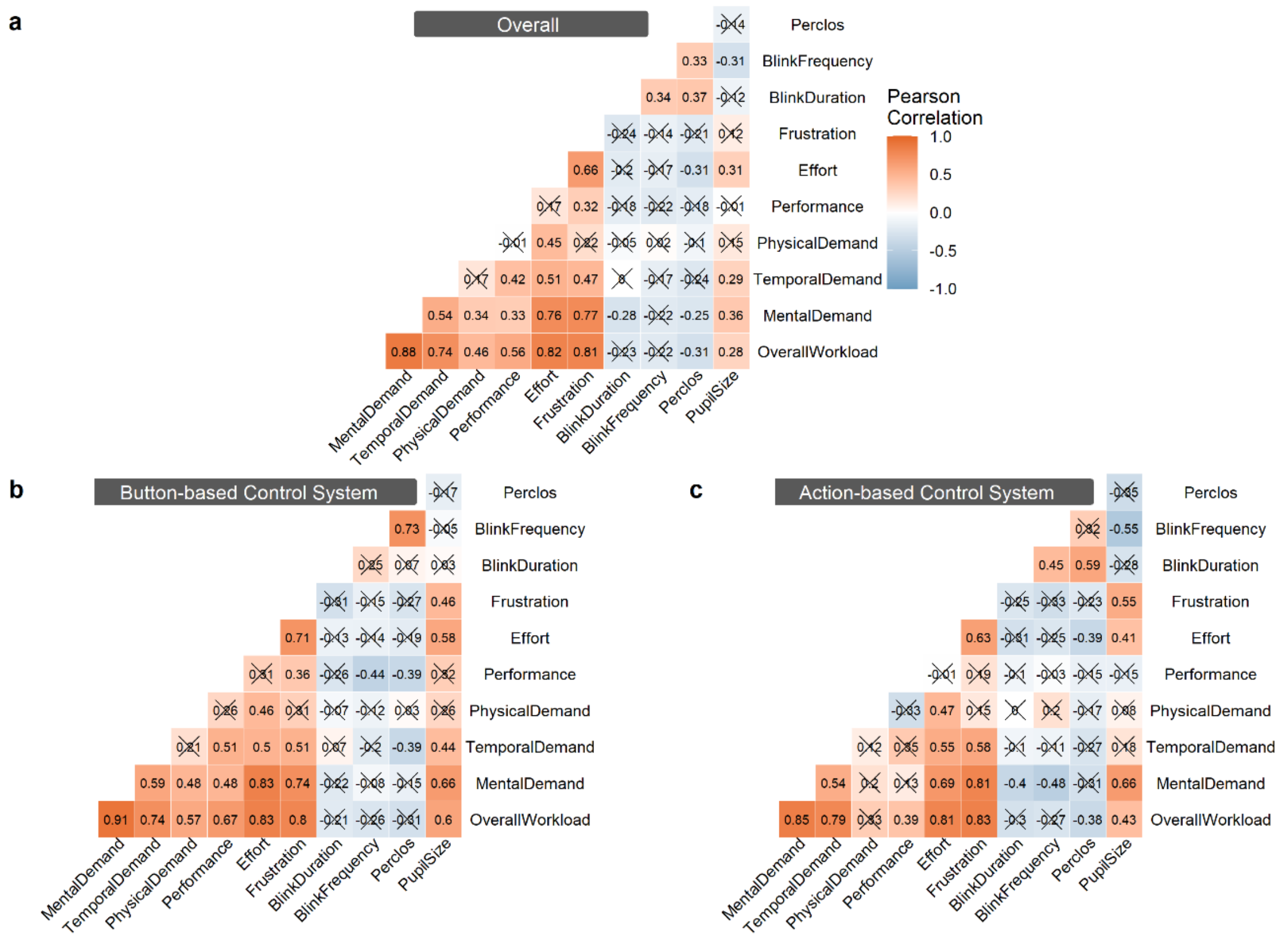

**Figure 10**. Correlation matrixes between NASA-TLX scores (Overall workload, Mental demand, Temporal demand, Physical demand, Performance, Effort, Frustration) and eye-tracking parameters (Blink duration, Blink frequency, Perclos, Pupil size). Figure a) depicts correlations on the overall dataset, independently from the control system used to teleoperate the robot; figure b) shows correlations within the button-based condition; figure c) shows correlations for the action-based control system.

# 5. DISCUSSION

Compared to traditional teleoperation means (e.g., desktop, mouse, keyboard), immersive VR devices allow users to physically act in the virtual surroundings and, potentially, to use their own body movements to guide robots and machines (e.g., Martin Barrio et al., 2019). Such feature would basically reproduce the manual guidance feature of collaborative robots, which allows humans to use their own hands to drive a robotic system over desired position. In this way, the amount of time typically spent on the physical robot programming is significantly streamlined and reduced (Faccio et al., 2022). However, little is known about how an action- rather than button-based control system of a virtual industrial robot impact users' performance and workload. This is due to the reasonable novelty of VR-based robotic teleoperations, but also to the lack of multidisciplinary and human factor-based investigations in telerobotics.

We thus built upon these opportunities and here proposed a systematic user-centered investigation of performance and workload during simulated robotic teleoperation in VR. The robot e-Series UR5e was faithfully virtualized and our participants guided it through a pick-and-place task via button-based and action-based control systems. The same task was also performed under low (single-task) and high (dual-task) mental demand. We leveraged all interaction data deriving from the tested human-robot collaborative framework, as well as eye-tracking data directly gained from the VR device. In this way, it was possible to monitor users' mental workload without interfering with the task. Furthermore, we collected self-reported workload and preferences for either control systems before and after the experiment. All in all, we draw a broad overview of the human psychophysical state by coupling workload and performance measurements throughout the tasks. Our main questions are deeply discussed in the following subparagraphs and included: the effectiveness of our task load manipulation in creating different levels of task demands (5.1); the effects of the enhanced interactivity of VR (action-based controls) compared to button-based control systems on human performance and workload (5.2); each eye parameter sensitivity to workload as recorded via the VR-embedded eye tracker throughout the tasks (5.3).

## 5.1 Task load

Our results are in line with our hypotheses and support the idea of a cognitive interference between the arithmetic task and the pick-and-place task. Specifically, performance and self-reported workload demonstrated the higher demand of the dual-tasking, while eye-parameters only partially reflected task load-related differences. According to the observed changes in perclos, our participants further demonstrated higher levels of vigilance in the dual-task with respect to the single-task. Interestingly, a tendency for increasing fatigue throughout the task performance can also be inferred by perclos and blink frequency variations. In the following subparagraphs (5.1.1, 5.1.2, 5.1.3), we thus unfold the effects of our task load manipulation on each of the investigated measures.

### 5.1.1 Performance measures

From a behavioral perspective, when performing the pick action concurrently with the arithmetic task, participants were slower and committed more errors compared to performing the pick action as a single task (Figure 4). Similarly, in the dual-task condition, they were more error-prone at the arithmetic task and it took longer for them to finalize the arithmetic sums compared to performing mental operations without additional tasks. Interestingly, the button-based condition revealed stronger dual-task behavioral effects, suggesting that it likely imposed higher demand compared to the action-based condition.

#### 5.1.2 Self-reports

Results from the NASA-TLX were on the same line as the users' performance (Figure 8). Participants reported significantly higher levels of workload when executing the pick-and-place task concurrently with the arithmetic task, indicating that they subjectively perceived higher demand compared to the single-task. Specifically, in the face of a comparable perceived physical demand, dual-tasking was reported as more mentally and temporally demanding, more frustrating, and more effortful compared to the single pick-and-place. Furthermore, participants self-rated their performance as significantly better in the single-task compared to the dual-task.

#### 5.1.3 Eye-tracking measures

The analysis of the eye parameters partially reflected the self-reported workload. As depicted in Figure 5, pupil size increased throughout the arithmetic task when moving from the start to the following arithmetic sums, no matter the control system involved. This indicated that our arithmetic task always imposed an increasing workload on the user. Furthermore, pupil size variation during the pick-and-place was significantly higher in the dual-task compared to the single-task, independently of the control system deployed. This finding is thus fairly robust and is in line with literature on teleoperation and/or robotics reporting higher pupil size for higher task load (Nenna et al., 2022; Wu et al., 2020; Zheng et al., 2015). Differently, perclos and blink parameters did not capture task load differences with the same accuracy as pupil size variations. Indeed, perclos and blink frequency were affected by the task load manipulation only when participants guided the robot through the action-based control system. One of the reasons behind this result, as deeply discussed in paragraph 5.2, might be related to the different difficulties of use of the two control systems (i.e., button- and action-based). Specifically for the action-based condition, on a macro-level, a lower perclos and blink frequency were observed in the dual-task, likely suggesting higher levels of vigilance compared to the single-task (Marquart et al., 2015). Similar results on workload-related blink variations were also observed in previous teleoperation research (Zheng et al., 2012; Guo et al., 2021). A common assumption is that users are likely to inhibit eye closures to reduce the risk of missing salient information (Fogarty and Stern, 1989), and such interpretation seems to apply to the present task too. On a micro-level, instead, perclos values gradually increased throughout the single-task from 2.74% on average in the first trials, to 5.26% in the last trials, while they only increased from 0.63% to 1.32% in the Dual-task. Similarly, blink frequency also decreased from 4.65 blink/min on average in the first trials, to 2.11 blink/min in the last trials of the task. Even though it was not supported by a statistical significance, this trend might reflect changes in the level of fatigue (Marquart et al., 2015): as time passed, users got tired, and their eye closures decreased. Another possible explanation is that performing the same monotonous task for some minutes can be tiring, thus affecting the level of vigilance in the task course (Körber et al., 2015).

### 5.2 Control system

Our findings confirmed the hypothesis of better performance and lower workload when using the action-based compared to the button-based control system in VR. Participants also demonstrated higher levels of vigilance throughout the whole pick-and-place executed via the button-based compared to the action-based control system. Performance, self-reports and eye-tracking measures differences between the two conditions were prominent, and they are deeply discussed in the following subparagraphs (5.2.1, 5.2.2, 5.2.3). Overall, this clear advantage of action-based controls might be related to embodied mechanisms involved in physical and direct operations, that can lead to more intuitive control of the virtual robot's movements in the 3D space. The hand-eye coordination is indeed a primal embodied behavior that makes every operation more affordable and natural. When guiding the robot via buttons, instead, the spatial intentions of the user need to be transposed from a 3-dimensional view to 4 static directions that run over two axes (forward-backward, left-right), which increases the operation complexity.

### 5.2.1 Performance measures

Participants were significantly faster when executing the teleoperations via physical action in both pick and place phases: they saved about 2 sec on average in each pick, and 1.5 sec in each place action compared to when using controller buttons. Furthermore, the error rate at the pick action decreased from about 17% to 7% when switching from controller buttons to physical actions, while the same advantage was not observed in the place action. Again, this is possibly due to the friendliness of the place operation, in which the diameter of the box where to place the bolt was much larger than the bolt itself. Furthermore, by looking at participants' performance at the arithmetic task, as compared to solely summing the presented numbers, the averaged error rate was almost doubled when additionally performing the pick-and-place task via physical actions, and even tripled when driving the robot via controller buttons. However, only the difference between error rates at the single arithmetic task and button-based dual-task was statistically significant. It thus seems that driving the robot via physical actions did not impose a degree of cognitive effort as high as when driving the same robot via controller buttons. In general, there is strong evidence in favor of using action-based controls for guiding the robotic arm. These findings are consistent with previous research that found better performance when using highly interactive control systems during robotic teleoperations (Vozar 2013; Franzluebbers and Johnsen, 2019; Gliesche et al., 2020; Martin-Barrio et al., 2020).

### 5.2.2 Self-reports

A generally higher perceived workload was observed in the button-based compared to the action-based condition. Furthermore, participants perceived higher frustration when driving the robot via buttons compared to physical actions. A tendency for higher mental demand was also observed in the button-based compared to the action-based condition, which however did not reach the significance threshold. These results are in accordance with the preference for either control system as rated after the whole experiment. Even before testing the teleoperation modalities, there was a clear tendency to prefer action-based control systems. This preference further increased after the experiment, whereby 89.47% of the tested sample reported personally preferring guiding the robot via physical actions, which was perceived as the less frustrating control system.

### 5.2.3 Eye-tracking measures

Notably, we did not intend to directly compare the control system-related eye parameters as they might be strongly influenced by the different movement magnitude involved in the action- and button-based conditions. This precaution was corroborated, for example, by the findings on pupil size variation: while in the button-based condition pupil size gradually increased from the start (window 1) to the end (window 6) of each action, in the action-based condition we observed a more rapid pupil size increase that reached its peak in windows 3 or 4, and then decreased (Figure 6). This could be related either to the larger physical motion involved in the action- compared to the button-based condition, which may have elicited higher arousal and activation, or to a constantly higher level of vigilance throughout the whole task session in the button-based condition, which may have flattened the pupil size variation. The latter interpretation seems to be further supported by the higher self-reported workload (Figure 8) and by the constantly lower level of Perclos observed in the button-based compared to action-based condition (Figure 7), which is known to be related to higher level of vigilance (Marquart et al., 2015).

On the same line of interpretation, the perclos difference between single- and dual-task was more evident in the action- than in the button-based condition, likely reflecting that executing the pick-and-place single-task via physical actions was so easy that it required very low vigilance compared to executing the same task via controller buttons. Furthermore, we noticed how perclos and blink duration were affected by our task load manipulation only in the action-based, but not in the button-based condition. Furthermore, when using the button-based control system, the task load manipulation similarly affected both pick and place actions, while when using the action-based control system, it only affected the pick action. Taken together, it seems that, no matter the task difficulty, participants were always investing more mental resources when operating via controller buttons rather than physical actions. This might have prevented the emergence of different blink and perclos trends between single- and dual-task, as well as a different pupil size trend between the pick and the place actions specifically in the button-based condition.

### 5.3 Sensitivity of VR-embedded eye-tracker metrics to workload

As discussed in paragraph 5.1, our task load manipulation was effective in producing two distinct levels of workload. From this observation, we further assessed the sensitivity of each eye-tracking metric to workload variations by correlating the self-reported workload and each eye-tracking parameter. From a first glimpse at Figure 10, independently from the control system involved, pupil size and perclos are particularly sensitive to changes in mental demand and effort, while blink duration responds specifically to mental demand. However, the latter relation was not supported by the results shown in Figure 7, as blink duration did not differ significantly between single- and dual-tasks. Differently, blink frequency did not show significant correlations with any workload dimensions on the overall task, suggesting that it might not be the best indicator of workload in VR. Furthermore, all observed relations in the overall task did not exceed the $R=0.36$, thus being quite weak.

When addressing the two control systems independently, the positive but weak relations between workload and pupil size observed in the overall task became even stronger. Relations between pupil size and mental demand reached a correlation of $R=0.66$ in each control system condition, almost doubling the R coefficient observed in the overall task. A positive relation between frustration and pupil size, which was not observed in the overall task, additionally stood in both control system conditions. This could be explained by the strong positive relations between self-reported mental demand and frustration ($R = 0.77$): the higher the mental demand, the higher the frustration, and the larger the pupil variation. Furthermore, it is worth briefly commenting on Figure 5, which shows trends in pupil size variations during the arithmetic task. Specifically, when the dual-task was performed via physical actions, there was higher variability in the pupil size trend compared to the button-based and to the arithmetic task conditions, in which the pupil increase was more linear throughout the task. Yet, a pupil size increase was captured across all conditions. Again, this is indicative of the resilience of such a metric in measuring workload under either higher (action-based) and lower (button-based) degrees of physical motion in VR. These findings on workload and pupil size are in line with robotics (Wu et al., 2020; Zheng et al., 2015) and virtual robotics research (Nenna et al., 2022).

While relations between pupil size and workload persist to the varied control system conditions, perclos and blink frequency better respond to workload fluctuations when using the action-based rather than the button-based control system. While only some weak relations were observed with temporal demand and performance when considering button-based actions, perclos showed stronger relations with overall workload and effort during action-based operations. Generally speaking, this is in line with the literature (Marquart et al., 2015) demonstrating how perclos is responsive to levels of vigilance and fatigue (which might be reflected in the 'effort' dimension in the NASA-TLX), but it is also in contrast with previous research in robotics that did not demonstrate significant relations between perclos and workload (Wu et al., 2020). Furthermore, if blink frequency did not yield any significant relation with workload in the overall task, it showed a moderate relation with mental demand exclusively in the action-based condition. This finding aligns with literature showing inverted relations between blink frequency and mental demand (Zheng et al., 2012; Borghini et al. 2014). Overall, the sensitivity of pupil size to workload stands out compared to the other eye metrics, which is consistent with previous research (Novak et al., 2015).

# 6. CONCLUSIONS, LIMITATIONS AND FUTURE WORKS

This work provide evidence that the VR-integrated eye-tracker can be particularly useful for continuous workload monitoring without interrupting ongoing tasks, thus matching precisely the human-centric digitalization wave promoted by Industry 4.0 and 5.0. Among the eye parameters investigated here, pupil size seems to be the most resilient indicator of workload, as it is not affected by the user's degree of physical motion. Furthermore, our results pointed towards action-based control systems as more efficient, intuitive, and less demanding solutions for teleoperating robots in VR. Such findings were gained through a human-centered approach. To bridge the lack of human-factors based research in the traditionally technical and engineering-based telerobotics research, we indeed deployed advanced digitalized technologies, while mainly focusing on the examination of human performance and workload rather than on the feasibility of the technical framework.

Nonetheless, we recognize the following limitations. First, real-world teleoperation tasks require more complex and varied activities. In our study, the choice of a simple experimental task such as the pick-and-place was intentional to guarantee appropriate experimental control yet allowing a natural behavior with the least possible constraints. Following studies might attempt to design more complex and diversified teleoperations. Second, due to the health emergency spread during the data collection, our results were gathered from a sample of young users (mainly students) with no prior experience with robot teleoperations. Future research might delve into this aspect by including participants that have more familiarity with teleoperation tasks for better understanding whether these findings transpose to teleoperation experts as well. Furthermore, it is to mention that the medium age of the European labor force in 2019 was about 40 (Statista, 2022), while the averaged age of our sample was about 26. On this point, it would be interesting to assess whether the same performance and workload trends observed in our young sample also apply to an older population, which is more likely representative of the eventual final users of such a technology. This would also help in understanding to what extent older users, who might be particularly unfamiliar with unconventional technologies such as VR, could be willing to accept such devices in their work life. Finally, we underline that the virtual robot used in the experimental study was not directly linked to the e-Series UR5e, and as such, was just a simulated teleoperation. This simulation was sufficient for fulfilling our aim of focusing on human performance and workload while driving an industrial robot in VR. However, we acknowledge that an actual link between the physical and the virtual model of the robot would constitute a more practical instance of telerobotics, also allowing to test technical aspects such as data streaming and effects of data transmission latency or disruptions. We are willing to complement our human-based framework with technical and engineering-based research in future works.

At a glance, this study paves the way for new perspectives in the telerobotics sector, which see eye-tracking-equipped VR as a valued resource in the ongoing 4.0 and 5.0 industrial revolutions. Such devices allow natural and embodied control of robotic systems, embracing the advantages of collaborative robotics in virtual spaces. We believe that such research line, which complement the innovations of telerobotics with knowledge coming from human factors and cognitive ergonomics sectors, will streamline and improve interactions between humans and robots, thus bringing substantial contribution to industry and society.

# REFERENCES

Ahlstrom, U., & Friedman-Berg, F. J. (2006). Using eye movement activity as a correlate of cognitive workload. International journal of industrial ergonomics, 36(7), 623-636.

Benedetto, S., Pedrotti, M., Minin, L., Baccino, T., Re, A., & Montanari, R. (2011). Driver workload and eye blink duration. *Transportation research part F: traffic psychology and behaviour*, *14*(3), 199-208.

Berkley, J. J. (2003). Haptic devices. *White Paper by Mimic Technologies Inc*, 1-4.

Borghini, G., Astolfi, L., Vecchiato, G., Mattia, D., & Babiloni, F. (2014). Measuring neurophysiological signals in aircraft pilots and car drivers for the assessment of mental workload, fatigue and drowsiness. Neuroscience & Biobehavioral Reviews, 44, 58-75.

Carswell, C. M., Clarke, D., & Seales, W. B. (2005). Assessing mental workload during laparoscopic surgery. *Surgical innovation*, *12*(1), 80-90.

Chien, S. Y., Lin, Y. L., Lee, P. J., Han, S., Lewis, M., & Sycara, K. (2018). Attention allocation for human multi-robot control: Cognitive analysis based on behavior data and hidden states. *International Journal of Human-Computer Studies*, *117*, 30-44.

Cicirelli, G., Attolico, C., Guaragnella, C., & D'Orazio, T. (2015). A kinect-based gesture recognition approach for a natural human robot interface. *International Journal of Advanced Robotic Systems*, *12*(3), 22.

Crespo, R., García, R., & Quiroz, S. (2015). Virtual reality application for simulation and off-line programming of the mitsubishi movemaster RV-M1 robot integrated with the oculus rift to improve students training. *Procedia Computer Science*, *75*, 107-112.

Dehais, F., Karwowski, W., & Ayaz, H. (2020). Brain at work and in everyday life as the next frontier: grand field challenges for neuroergonomics. *Frontiers in Neuroergonomics*, 1.

Delignette-Muller, M. L., & Dutang, C. (2015). fitdistrplus: An R package for fitting distributions. *Journal of statistical software*, *64*, 1-34.

Doolani, S., Wessels, C., Kanal, V., Sevastopoulos, C., Jaiswal, A., Nambiappan, H., & Makedon, F. (2020). A review of extended reality (xr) technologies for manufacturing training. *Technologies*, *8*(4), 77.

Du, G., Zhang, L., Su, K., Wang, X., Teng, S., & Liu, P. X. (2022). A Multimodal Fusion Fatigue Driving Detection Method Based on Heart Rate and PERCLOS. *IEEE Transactions on Intelligent Transportation Systems*.

Faccio, M., Granata, I., Menini, A., Milanese, M., Rossato, C., Bottin, M., ... & Rosati, G. (2022). Human factors in cobot era: a review of modern production systems features. *Journal of Intelligent Manufacturing*, 1-22.

Faure, V., Lobjois, R., & Benguigui, N. (2016). The effects of driving environment complexity and dual tasking on drivers' mental workload and eye blink behavior. *Transportation research part F: traffic psychology and behaviour*, *40*, 78-90.

Fogarty, C., & Stern, J. A. (1989). Eye movements and blinks: their relationship to higher cognitive processes. *International journal of psychophysiology*, *8*(1), 35-42.

Franzluebbers, A., & Johnson, K. (2019, October). Remote robotic arm teleoperation through virtual reality. In *Symposium on Spatial User Interaction* (pp. 1-2).

Grandi, F., Zanni, L., Peruzzini, M., Pellicciari, M., & Campanella, C. E. (2020). A Transdisciplinary digital approach for tractor's human-centred design. *International Journal of Computer Integrated Manufacturing*, *33*(4), 377-397.

Gliesche, P., Krick, T., Pfingsthorn, M., Drolshagen, S., Kowalski, C., & Hein, A. (2020). Kinesthetic Device vs. Keyboard/Mouse: A Comparison in Home Care Telemanipulation. *Frontiers in Robotics and AI*, 172.

Grabowski, A., Jankowski, J., & Wodzyński, M. (2021). Teleoperated mobile robot with two arms: the influence of a human-machine interface, VR training and operator age. *International Journal of Human-Computer Studies*, *156*, 102707.

Guo, Y., Freer, D., Deligianni, F., & Yang, G. Z. (2021). Eye-tracking for performance evaluation and workload estimation in space telerobotic training. *IEEE Transactions on Human-Machine Systems*, *52*(1), 1-11.

Hart, S. G., & Staveland, L. E. (1988). Development of NASA-TLX (Task Load Index): Results of empirical and theoretical research. In *Advances in psychology* (Vol. 52, pp. 139-183). North-Holland.

Holland, M. K., & Tarlow, G. (1972). Blinking and mental load. *Psychological Reports*, *31*(1), 119-127.

*ISO 9241-210:2019*. (n.d.). ISO. https://www.iso.org/standard/77520.html

Kahneman, D. (1973). *Attention and effort* (Vol. 1063, pp. 218-226). Englewood Cliffs, NJ: Prentice-Hall.

Kim, S. Y., Park, H., Kim, H., Kim, J., & Seo, K. (2022). Technostress causes cognitive overload in high-stress people: Eye tracking analysis in a virtual kiosk test. *Information Processing & Management*, *59*(6), 103093.

Köles, M. (2017). A review of pupillometry for human-computer interaction studies. Periodica Polytechnica Electrical Engineering and Computer Science, 61(4), 320-326

Körber, M., Cingel, A., Zimmermann, M., & Bengler, K. (2015). Vigilance decrement and passive fatigue caused by monotony in automated driving. *Procedia Manufacturing*, *3*, 2403-2409.

Krüger, J., Lien, T. K., & Verl, A. (2009). Cooperation of human and machines in assembly lines. *CIRP annals*, *58*(2), 628-646.

Lu, Y., Zheng, H., Chand, S., Xia, W., Liu, Z., Xu, X., ... & Bao, J. (2022). Outlook on human-centric manufacturing towards Industry 5.0. *Journal of Manufacturing Systems*, *62*, 612-627.

Marinescu, A. C., Sharples, S., Ritchie, A. C., Sanchez Lopez, T., McDowell, M., & Morvan, H. P. (2018). Physiological parameter response to variation of mental workload. Human factors, 60(1), 31-56.

Martin, S., & Hillier, N. (2009, December). Characterisation of the Novint Falcon haptic device for application as a robot manipulator. In *Australasian Conference on Robotics and Automation (ACRA)* (pp. 291-292). Citeseer.

Martín-Barrio, A., Roldán, J. J., Terrile, S., del Cerro, J., & Barrientos, A. (2020). Application of immersive technologies and natural language to hyper-redundant robot teleoperation. *Virtual Reality*, *24*(3), 541-555.

Marquart, G., Cabrall, C., & de Winter, J. (2015). Review of eye-related measures of drivers' mental workload. Procedia Manufacturing, 3, 2854-2861

Matthews, G., Reinerman-Jones, L. E., Barber, D. J., & Abich IV, J. (2015). The psychometrics of mental workload: Multiple measures are sensitive but divergent. *Human factors*, *57*(1), 125-143.

Mavridis, N., Pierris, G., Gallina, P., Moustakas, N., & Astaras, A. (2015, July). Subjective difficulty and indicators of performance of joystick-based robot arm teleoperation with auditory feedback. In *2015 International Conference on Advanced Robotics (ICAR)* (pp. 91-98). IEEE.

McIntire, L. K., McKinley, R. A., Goodyear, C., & McIntire, J. P. (2014). Detection of vigilance performance using eye blinks. *Applied ergonomics*, *45*(2), 354-362.

Naranjo, J. E., Sanchez, D. G., Robalino-Lopez, A., Robalino-Lopez, P., Alarcon-Ortiz, A., & Garcia, M. V. (2020). A scoping review on virtual reality-based industrial training. *Applied Sciences*, *10*(22), 8224.

Nenna, F., Orso, V., Zanardi, D., & Gamberini, L. (2022). The virtualization of human–robot interactions: a user-centric workload assessment. *Virtual Reality*, 1-19.

Novak, D., Beyeler, B., Omlin, X., & Riener, R. (2015). Workload estimation in physical human–robot interaction using physiological measurements. *Interacting with computers*, *27*(6), 616-629.

Pashler, H. (1984). Processing stages in overlapping tasks: evidence for a central bottleneck. *Journal of Experimental Psychology: Human perception and performance*, *10*(3), 358.

Pomplun, M., & Sunkara, S. (2019). Pupil dilation as an indicator of cognitive workload in human-computer interaction. In *Human-Centered Computing* (pp. 542-546). CRC Press.

Rebelo, F., Noriega, P., Duarte, E., & Soares, M. (2012). Using virtual reality to assess user experience. *Human factors*, *54*(6), 964-982.

Rosen, E., Whitney, D., Phillips, E., Ullman, D., & Tellex, S. (2018, March). Testing robot teleoperation using a virtual reality interface with ROS reality. In Proceedings of the 1st International Workshop on Virtual, Augmented, and Mixed Reality for HRI (VAM-HRI) (pp. 1-4).

Rouanet, P., Béchu, J., & Oudeyer, P. Y. (2009, September). A comparison of three interfaces using handheld devices to intuitively drive and show objects to a social robot: the impact of underlying metaphors. In *RO-MAN 2009-The 18th IEEE International Symposium on Robot and Human Interactive Communication* (pp. 1066-1072). IEEE.

Smith, K., & Sepasgozar, S. (2022). Governance, Standards and Regulation: What Construction and Mining Need to Commit to Industry 4.0. *Buildings*, *12*(7), 1064.

Statista. (2022, August 5). *Median age of the global labor force by region and gender 2019*. Retrieved September 27, 2022, from https://www.statista.com/statistics/996588/median-age-global-labor-force-region-gender/

Villani, V., Righi, M., Sabattini, L., & Secchi, C. (2020). Wearable devices for the assessment of cognitive effort for human–robot interaction. *IEEE Sensors Journal*, *20*(21), 13047-13056.

Vozar, S. E. (2013). *A Framework for Improving the Speed and Performance of Teleoperated Mobile Manipulators* (Doctoral dissertation).

Wang, X. V., Kemény, Z., Váncza, J., & Wang, L. (2017). Human–robot collaborative assembly in cyber-physical production: Classification framework and implementation. *CIRP annals*, *66*(1), 5-8.

Wu, C., Cha, J., Sulek, J., Zhou, T., Sundaram, C. P., Wachs, J., & Yu, D. (2020). Eye-tracking metrics predict perceived workload in robotic surgical skills training. *Human factors*, *62*(8), 1365-1386.

You, E., & Hauser, K. (2012, June). Assisted teleoperation strategies for aggressively controlling a robot arm with 2d input. In *Robotics: science and systems* (Vol. 7, p. 354). USA: MIT Press.

Zheng, B., Jiang, X., Tien, G., Meneghetti, A., Panton, O. N. M., & Atkins, M. S. (2012). Workload assessment of surgeons: correlation between NASA TLX and blinks. Surgical endoscopy, 26(10), 2746-2750.

Zheng, B., Jiang, X., & Atkins, M. S. (2015). Detection of changes in surgical difficulty: evidence from pupil responses. *Surgical innovation*, *22*(6), 629-635.